\documentclass[groupedaddress,preprint]{revtex4-1}
\usepackage{graphicx}
\usepackage{amsmath}

\begin{document}

\title{Two-photon imaging assisted by a dynamic random medium}

\author{Peilong Hong}\email{plhong@njust.edu.cn}
\affiliation{School of Science, Nanjing University of Science and Technology, Nanjing, Jiangsu 210094, China}

\date{\today}

\begin{abstract}
Random scattering is usually viewed as a serious nuisance in optical imaging, and needs to be prevented in the conventional imaging scheme based on single-photon interference.
Here we proposed a two-photon imaging scheme with the widely used lens replaced by a dynamic random medium. In contrast to destroying imaging process, the dynamic random medium in our scheme works as a crucial imaging element to bring constructive interference, and allows us to image an object from light field scattered by this dynamic random medium.
On the one hand, our imaging scheme with incoherent two-photon illumination enables us to achieve super-resolution imaging with the resolution reaching Heisenberg limit.
On the other hand, with coherent two-photon illumination, the image of a pure-phase object can be obtained in our imaging scheme.
These results show new possibilities to overcome bottleneck of widely used single-photon imaging by developing imaging method based on multi-photon interference.
\end{abstract}


\maketitle


Optical imaging is a powerful tool in many fields, ranging from frontier scientific research to our everyday life. The most fundamental and widely used imaging scheme is a lens-assisted imaging scheme, which requires finely designed lens to introduce multi-path constructive interference for imaging~\cite{brooker2003modern}. Scattering by random inhomogeneities is extremely harmful for such lens-assisted imaging scheme, since random scattering seriously distorts the wavefront such that constructive interference essential for optical imaging is destroyed~\cite{goodman2000statistical}. However, random scattering is inevitable in many imaging scenarios, such as imaging through a rough wall, imaging through turbulent air, and biologic imaging through organic tissues.
Phase conjugation (or time reversal) is a well developed method to overcome the drawback of random scattering by compensating the wavefront distortion~\cite{he2002optical,mosk2012controlling}.
Recent advances show that one can obtain the intensity autocorrelation of an object from the scattering light in properly arranged configurations, from which an image of the object can be extracted through proper iterative algorithms~\cite{bertolotti2012non,katz2014non}.
In our work, we propose a two-photon imaging scheme, in which a dynamic random medium, instead of destroying imaging process, plays a key role for directly imaging an object.

Two-photon interference effect was first reported by Hanbury Brown and Twiss in 1956~\cite{brown1956test}, and later generally formulated by Glauber~\cite{glauber1963quantum}. With different light sources such as thermal source, single-photon source, entangled two-photon source, and phase-controlled source, a lot of interesting two-photon interference effects have been reported, including bunching and superbunching effect of thermal light~\cite{brown1956test,hong2012two}, antibunching effect~\cite{kimble1977photon}, Hong-Ou-Mandel effect~\cite{hong1987measurement}, and subwavelength interference~\cite{fonseca1999measurement,boto2000quantum,edamatsu2002measurement,d2001two,scarcelli2004two,wang2004subwavelength,xiong2005experimental,hong2013subwavelength,hong2015super}, just to name a few.
Interestingly, it was found that optical imaging based on two-photon interference can be established if indistinguishable two-photon paths are properly designed to form constructive interference, leading to the observation of ghost imaging first by using an entangled two-photon source~\cite{pittman1995optical} and later by using a thermal source~\cite{gatti2004ghost,valencia2005two,ferri2005high}. In this work, we realize constructive interference of multiple two-photon paths by considering the specific two-photon coherence property of a dynamic random medium and that of an entangled two-photon source, such that a direct image of the object is obtained by measuring two-photon correlation of the scattered light field.

\vspace{12pt}
\noindent\textbf{\large{Results}}

\begin{figure}[!htb]
    \centering
    \includegraphics[width=0.8\textwidth]{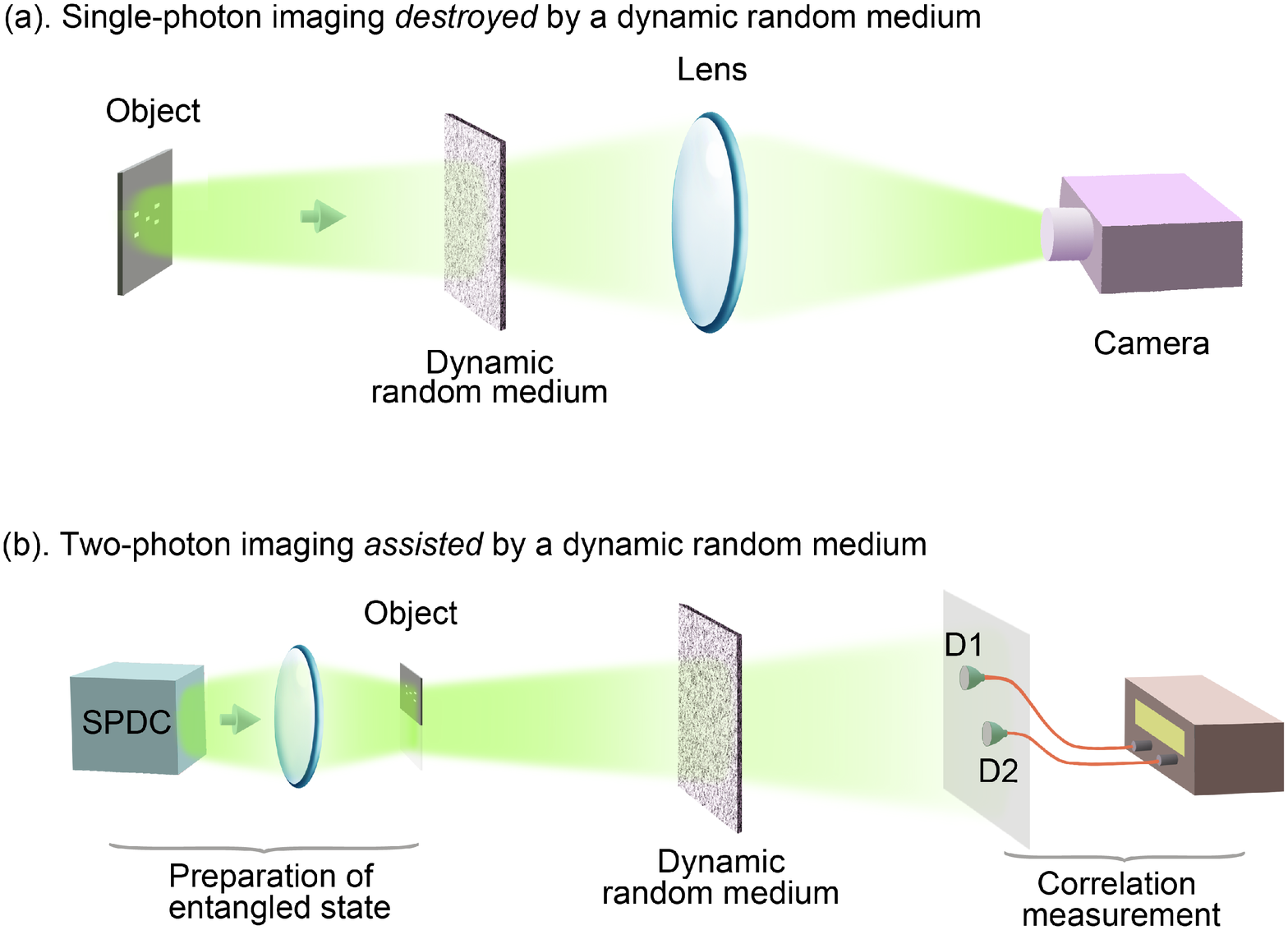}
    \caption{(a). Single-photon imaging with a lens. The imaging process is destroyed by a dynamic random media located between object and image plane.
    (b). Two-photon imaging assisted by a dynamic random medium. An entangled two-photon source (SPDC) is used as illumination, and the object is put at top half of the Fourier plane of the source. In the image plane, two-photon correlation measurement is employed to observe two-photon interference. \label{Fig1_scheme}}
\end{figure}

\vspace{6pt}\noindent\textbf{Imaging scheme.}
Figure.~\ref{Fig1_scheme}(b) shows our two-photon imaging scheme, where a dynamic random medium is located
between the object and image plane.
In this geometry, light field is first totally distorted by the random medium, then observed in the image plane.
This geometry makes the imaging process destroyed in a single-photon imaging scheme as shown in Fig.~\ref{Fig1_scheme}(a).
However, in our two-photon imaging scheme, the dynamic random medium works as a key element for directly imaging an object when the object is illuminated by an entangled two-photon source generated through spontaneous parametric down conversion process (SPDC)~\cite{shih2003entangled,edgar2012imaging}.

\begin{figure}[!htb]
    \centering
    \includegraphics[width=0.6\textwidth]{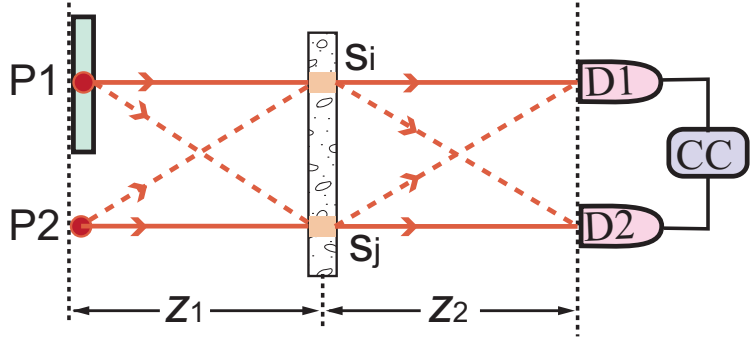}
    \caption{Different but indistinguishable two-photon paths for a pair of entangled photons triggering a coincidence count in the two-photon imaging scheme shown in Fig.~\ref{Fig1_scheme}(b). Here $z_1$ represents the distance between the object plane and the random medium plane, and $z_2$ is the distance between the random medium plane and the image plane. \label{Fig2_superposition}}
\end{figure}

\vspace{6pt}\noindent\textbf{Two-photon coherence property of a thin dynamic random medium.} We consider a thin dynamic random medium, and its aperture function is formulated as a complex random phasor ${\cal{R}}({\bf{\rho}}_r,t)$~\cite{goodman2000statistical}, where ${\bf{\rho}}_r=(x_r,y_r)$ is the spatial coordinate on the medium plane and $t$ is the temporal variable.
The first-order correlation function of the random medium is  $G^{(1,1)}({\bf{\rho}}_{r1};{\bf{\rho}}_{r2})=\langle{\cal{R}}^*({\bf{\rho}}_{r1}) {\cal{R}}({\bf{\rho}}_{r2})\rangle = \delta({\bf{\rho}}_{r1}-{\bf{\rho}}_{r2})$, where $\langle \cdots \rangle$ denotes the ensemble average, and $\delta(\cdots)$ is the Dirac delta function. We have dropped the temporal variable $t$ in the random phasor ${\cal{R}}({\bf{\rho}}_r,t)$ for simplicity. Therefore, the dynamic random medium is a first-order incoherent aperture which destroys single-photon interference.
In the domain of two-photon interference, second-order correlation property, rather than first-order correlation property, plays a key role~\cite{glauber1963quantum}.
The second-order correlation function of a thin dynamic random medium is~\cite{wang2004subwavelength,cheng2004incoherent,hong2015super}
\begin{equation} \label{Rand_medium}
\begin{split}
G_R^{(2,2)}({\bf{\rho}}_{r1},{\bf{\rho}}_{r2};{\bf{\rho}}_{r3},{\bf{\rho}}_{r4})
             &=\langle {\cal{R}}^*({\bf{\rho}}_{r1}){\cal{R}}^*({\bf{\rho}}_{r2}){\cal{R}}({\bf{\rho}}_{r3}){\cal{R}}({\bf{\rho}}_{r4})\rangle \\
              &= \delta({\bf{\rho}}_{r1}-{\bf{\rho}}_{r3})\delta({\bf{\rho}}_{r2}-{\bf{\rho}}_{r4})+\delta({\bf{\rho}}_{r1}-{\bf{\rho}}_{r4})\delta({\bf{\rho}}_{r2}-{\bf{\rho}}_{r3}),
\end{split}
\end{equation}
which leads to Hanbury Brown-Twiss type two-photon interference~\cite{agafonov2008high}.
To clearly understand this two-photon interference, let us analyze the indistinguishable two-photon paths originated from a dynamic random medium~\cite{hong2012two,scarcelli2004two,mandel1999quantum,nevet2011indistinguishable,fano1961quantum,agafonov2008high,hong2013subwavelength}. For a pair of photons from any two points of the dynamic random medium, there are always two alternative indistinguishable paths for the pair of photons to trigger a coincidence counts as shown in Fig~\ref{Fig2_superposition}, i.e., photon from point source $S_i$ goes to detector D1 while photon from point source $S_j$ goes to detector D2, and vice versa. Since a dynamic random medium consists of many secondary point sources, a dynamic random medium offers us a lot of two-photon interferometers, each of which introduces superposition of two indistinguishable two-photon paths. Consequently, two-photon paths originated from an object will be reshaped by the random medium, raising the posibility to estabilish two-photon imaging with the help of a dynamic random medium. Interestingly, we find that, by introducing entangled two-photon illumination, two-photon imaging process is naturally established without any further optical engineering.

\vspace{6pt}\noindent\textbf{Coherent and incoherent two-photon illumination.} In our imaging scheme, we consider an entangled two-photon source generated from type-I phase matched SPDC process~\cite{shih2003entangled,edgar2012imaging}. Because of the phase-matching, the pair of degenerate down-converted photons are momentum correlated. Therefore, by projecting the light from the source plane to its Fourier plane, the second-order correlation function evolves to be~\cite{shih2003entangled,edgar2012imaging,tasca2009propagation}
\begin{equation}\label{Entangle_Source_coherent}
G_{coh}^{(2;2)}({\bf{\rho}}_{01},{\bf{\rho}}_{02};{\bf{\rho}}_{03},{\bf{\rho}}_{04}) = \delta({\bf{\rho}}_{01}+{\bf{\rho}}_{02})\delta({\bf{\rho}}_{03}+{\bf{\rho}}_{04}) \,,
\end{equation}
where ${\bf{\rho}}_{0i}=(x_{0i},y_{0i}) (i=1,2,3,4)$ is the coordinate in the Fourier plane, and we assume the ensemble-averaged light intensity in the Fourier plane is homogeneous for simplicity. From Eq.~(\ref{Entangle_Source_coherent}), one sees that the emitted pair of photons have a position entanglement in the Fourier plane, i.e., when one photon is found at one position ${\bf{\rho}}_{0n}$, its twin photon will always be found at corresponding central symmetric position $-{\bf{\rho}}_{0n}$ with respect to the center position ${\bf{\rho}}_0={\bf{0}}$. We put the object in the top quadrant ($y_0\geq0$) of the Fourier plane, and let the bottom quadrant ($y_0<0$) be transmitted as shown in Fig.~\ref{Fig1_scheme}(b).

Without any post-manipulation, the entangled two-photon source is two-photon coherent one, since different pairs of correlated position $({\bf{\rho}}_{0n}, -{\bf{\rho}}_{0n})$ is two-photon coherent as shown in Eq.~(\ref{Entangle_Source_coherent}). The coherent source can be tuned to be an incoherent one by inserting a dynamic random-phase screen such as a rotating ground glass disk on the Fourier plane (or accompanied transmitted part of the Fourier plane), so that two-photon interference between different pairs of position $({\bf{\rho}}_{0n}, -{\bf{\rho}}_{0n})$ is erased. The second-order correlation function in the incoherent case is thus expressed as
\begin{equation}\label{Entangle_Source_incoherent}
G_{inc}^{(2;2)}({\bf{\rho}}_{01},{\bf{\rho}}_{02};{\bf{\rho}}_{03},{\bf{\rho}}_{04}) = \delta({\bf{\rho}}_{01}+{\bf{\rho}}_{02})\delta({\bf{\rho}}_{03}+{\bf{\rho}}_{04})
 \cdot [\delta({\bf{\rho}}_{01}-{\bf{\rho}}_{03})+\delta({\bf{\rho}}_{01}+{\bf{\rho}}_{03})].
\end{equation}

\vspace{6pt}\noindent\textbf{Superposition of multiple two-photon paths.} To understand the two-photon imaging process, let us analyze superposition of two-photon paths in our imaging scheme. Figure~\ref{Fig2_superposition} shows that, for a pair of entangled photons $P1$ and $P2$ at the object plane, there are four alternative indistinguishable two-photon paths for them to trigger a coincidence count, i.e., $(P1-S_i-D1:P2-S_j-D2)$, $(P1-S_i-D2:P2-S_j-D1)$, $(P1-S_j-D1:P2-S_i-D2)$ and $(P1-S_j-D2:P2-S_i-D1)$.
Note that, when photon $P1$ transmits through the object at position ${\bf{\rho}}_{0n}$, the other photon $P2$ always comes from the central symmetric point $-{\bf{\rho}}_{0n}$ in the accompanied transmitted plane due to the position correlation of the entangled biphoton.
In a specific situation that $z_1=z_2$, when $D_1$ and $D_2$ are at symmetric positions of $P1$ and $P2$ regarding the random medium plane respectively, the two paths $(P1-S_i-D2:P2-S_j-D1)$ and $(P1-S_j-D2:P2-S_i-D1)$ are always in phase for different pairs of $(S_i, S_j)$ shown in Fig.~\ref{Fig2_superposition}. Constructive interference of these in-phase two-photon paths directly leads to an image of the object in the case of incoherent two-photon illumination. In the case of coherent two-photon illumination, more indistinguishable two-photon paths need to be considered because different groups of two-photon paths from different pairs $({\bf{\rho}}_{0n},-{\bf{\rho}}_{0n})$ are also indistinguishable. Consequently, there are cross interference between different parts of the object, which will lead to more features in the observed interference pattern.

In order to clearly show the general imaging conditions and image patterns in both the incoherent and coherent illumination situations, we will theoretically calculate the two-photon correlation function in the image plane in the following.


\vspace{6pt}\noindent \textbf{Theoretical derivation of imaging condition.} By taking the paraxial approximation for a quasi-monochromatic light propagating in free space, the light field operator in the image plane is expressed as~\cite{valencia2005two,brooker2003modern,hong2012two,ferri2005high}
\begin{equation}
E^{(+)}({\bf{\rho}}) \propto \int E_0^{(+)}({\bf{\rho}}_0) A({\bf{\rho}}_0) Q({\bf{\rho}}_r-{\bf{\rho}}_0,z_1)    {\cal{R}}({\bf{\rho}}_r)
 Q({\bf{\rho}}-{\bf{\rho}}_r,z_2)  d{\bf{\rho}}_0 d{\bf{\rho}}_r \,,
\end{equation}
where ${\bf{\rho}}_0$, ${\bf{\rho}}_r$, and ${\bf{\rho}}$ are the transverse coordinates in the object plane, the random medium plane, and the image plane, respectively. $E_0^{(+)}({\bf{\rho}}_0)$ is the field operator of entangled two-photon light that illuminates the object.
$ A({\bf{\rho}}_0) =O({\bf{\rho}}_0) (y_0\geq0) + 1 (y_0<0) $, which is the aperture function of the object plane. $O({\bf{\rho}}_0) (y \geq0)$ is the object aperture function, and the complex amplitude of associated transmitted plane is assumed to be unity.
$Q({\bf{\rho}}_p-{\bf{\rho}}_q,z_i)  = \exp (ik|{\bf{\rho}}_p-{\bf{\rho}}_q|^2/(2z_i)) $ is the quadratic phase factor introduced by the light diffraction in free space, where $k$ is the wave vector, and $z_i$ is the distance between ${\bf{\rho}}_p$ plane and ${\bf{\rho}}_q$ plane.
Note that, any single-photon interference pattern in the image plane is erased, since single-photon interference from different regions of the random scattering medium is totally destroyed.

The second-order correlation function between two positions ${\bf{\rho}}_1$ and ${\bf{\rho}}_2$ in the image plane is expressed as~\cite{glauber1963quantum,hong2012two}
\begin{equation}
\begin{split}
G^{(2)}({\bf{\rho}}_1;{\bf{\rho}}_2)   &=\langle E^{(-)}({\bf{\rho}}_1)E^{(-)}({\bf{\rho}}_2)E^{(+)}({\bf{\rho}}_2)E^{(+)}({\bf{\rho}}_1) \rangle \\
                          \propto & \int  G^{(2;2)}({\bf{\rho}}_{01},{\bf{\rho}}_{02};{\bf{\rho}}_{03},{\bf{\rho}}_{04})
                           A^*({\bf{\rho}}_{01}) A^*({\bf{\rho}}_{02}) A({\bf{\rho}}_{03}) A({\bf{\rho}}_{04})  \\
                          & \times Q^*({\bf{\rho}}_{r1}-{\bf{\rho}}_{01}) Q^*({\bf{\rho}}_{r2}-{\bf{\rho}}_{02}) Q({\bf{\rho}}_{r3}-{\bf{\rho}}_{03})  Q({\bf{\rho}}_{r4}-{\bf{\rho}}_{04}) \\
                            &\, \times    G_R^{(2;2)}({\bf{\rho}}_{r1},{\bf{\rho}}_{r2};{\bf{\rho}}_{r3},{\bf{\rho}}_{r4})
                                    Q^*({\bf{\rho}}_{1}-{\bf{\rho}}_{r1}) Q^*({\bf{\rho}}_{2}-{\bf{\rho}}_{r2})     \\
                            &\, \times  Q({\bf{\rho}}_{2}-{\bf{\rho}}_{r3})  Q({\bf{\rho}}_{1}-{\bf{\rho}}_{r4})
                             \, d{\bf{\rho}}_{01} \cdots d{\bf{\rho}}_{04} \, d{\bf{\rho}}_{r1} \cdots d{\bf{\rho}}_{r4} \,,
\end{split}
\end{equation}
where we have dropped variable $z_i$ in the expression of quadratic phasor for simplicity. By using the results of Eqs.~(\ref{Rand_medium})-(\ref{Entangle_Source_incoherent}) and $A({\bf{\rho}}_{0}) = O({\bf{\rho}}_0) (y_0\geq0) + 1 (y_0<0)$, we derive to obtain the second-order correlation function in the case of

1). incoherent two-photon illumination
\begin{equation}\label{Image_1}
G_{img1}^{(2)}({\bf{\rho}}_1;{\bf{\rho}}_2)   \propto G_{bg1} + \int  \big| O({\bf{\rho}}_{0}) \big| ^2
   \text{somb}^2 \Big(\frac{kD}{2z_1}\big |{2\bf{\rho}}_{0} \pm \frac{\Delta{\bf{\rho}}_{21}}{m}\big| \Big) d{\bf{\rho}}_{0} \,,
\end{equation}
where the plus-minus sign means the total result is the sum of two possible results. $D$ is the diameter of the random medium, $\Delta{\bf{\rho}}_{21} = {\bf{\rho}}_2-{\bf{\rho}}_1$, and the magnification factor $m=z_2/z_1$. $\text{somb} (\cdots)$ is the sombrero function, which is known as point-spread function (PSF) in an imaging system~\cite{brooker2003modern}. $G_{bg1}=  2 \, \int  |O({\bf{\rho}}_{0})|^2 \big(1+
\text{somb}^2 (\frac{kD|{\bf{\rho}}_{0}|}{z_1})
 +  \text{somb}^2 (\frac{kD |\Delta{\bf{\rho}}_{21}| }{2z_2}) \big) d{\bf{\rho}}_{0}$, which only contributes as a background.
Besides of this background $G_{bg1}$, the other term of Eq.~(\ref{Image_1}) represents a point-to-point direct image of the object.

\begin{figure}[!htb]
    \centering
    \includegraphics[width=0.85\textwidth]{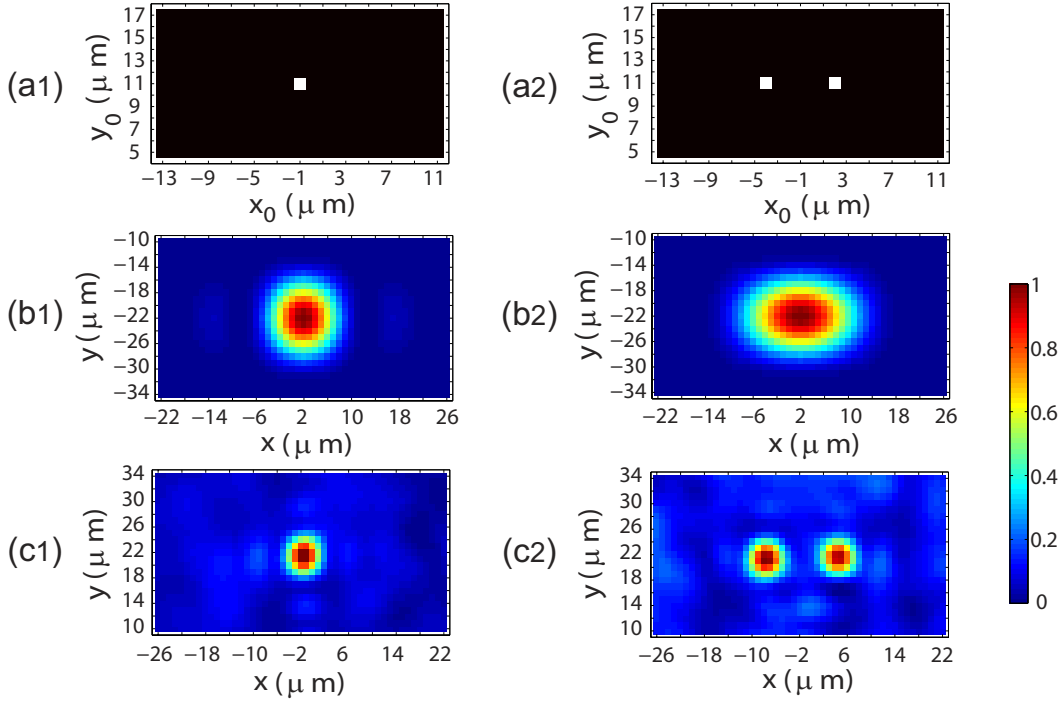}
    \caption{Numerical simulations of lens-assisted single photon imaging (b1) and (b2), and super-resolution two-photon imaging with incoherent two-photon illumination (c1) and (c2).
    The original two objects for imaging are shown in (a1) and (a2), respectively.
    Simulation parameters for two-photon imaging are $z_1=20$ cm, $z_2=2 \cdot z_1$, the coordinates of both detectors are set to be ${\bf{\rho}}_2=-{\bf{\rho}}_1={\bf{\rho}}$ with ${\bf{\rho}}=(x,y)$, the size of the dynamic scattering medium is 2 cm $\times$ 2 cm, and light wavelength $\lambda=532$ nm.
    To simulate the len-assisted single-photon imaging, we simply replaces the dynamic random medium with a lens, and set the lens to be the same size as that of the dynamic random medium. To meet the imaging condition, the focal length of the imaging lens is set to be $f=z_1\cdot z_2/(z_1+z_2)$.
    Therefore, the numerical aperture of the single-photon imaging scheme is the same as that of the two-photon imaging scheme.
    All plots are normalized so that the minimum and maximum value are 0 and 1, respectively. \label{Fig3_Super_resolution}}
\end{figure}

Interestingly, by comparing the PSF of Eq.~(\ref{Image_1}) with that of single-photon imaging system of the same numerical aperture $\text{NA}=D/(2z_1)$, we find that the resolution of our two-photon imaging system surpasses the Rayleigh resolution bound by a factor of 2~\cite{brooker2003modern,giovannetti2009sub}. This resolution enhancement is due to the fact that the effective wavelength of two-photon de Broglie wave is half of that of the light photon~\cite{fonseca1999measurement,boto2000quantum,edamatsu2002measurement,d2001two,scarcelli2004two,wang2004subwavelength,xiong2005experimental,hong2013subwavelength,hong2015super,giovannetti2009sub,jacobson1995photonic}.
To clearly show this super-resolution effect, we did numerical simulation (see Methods) by considering a system with parameters: $D=2$ cm, $z_1=20$ cm, $z_2=2\cdot z_1$, and wavelength $\lambda=532$ nm. Therefore, the numerical aperture is $NA=0.05$, and the magnification factor $m$ is 2. With these parameters, the Rayleigh resolution bound for single-photon imaging is 6.5 $\mu$m, while the resolution of our two-photon imaging system is 3.2 $\mu$m. Figures~\ref{Fig3_Super_resolution} (a1), (b1) and (c1) show the object (a single transmission square of size 1 $\mu$m $\times$ 1 $\mu$m), the image spot of the object in a lens-assisted (focal length $f=z_1\cdot z_2/(z_1+z_2)$) single-photon imaging scheme, and the image spot in our two-photon imaging scheme, respectively. One clearly sees that the size of the image spot in our two-photon imaging scheme is much smaller than that in a lens-assisted imaging scheme. To further confirm this resolution enhancement, we also simulated the imaging of an object that is two 1 $\mu$m $\times$ 1 $\mu$m transmission squares of distance 6 $\mu$m. Figures~\ref{Fig3_Super_resolution} (a2), (b2) and (c2) show the object, the image of the object in a lens-assisted single-photon imaging scheme, and the image of object in our two-photon imaging scheme, respectively. It is apparent that the traditional lens-assisted imaging can not distinguish the two squares of the object, since their distance (6 $\mu$m) is smaller than the Rayleigh resolution bound (6.5 $\mu$m). However, our two-photon imaging scheme has an enhanced resolution bound (3.2 $\mu$m), therefore, the two image spots in our two-photon imaging scheme are clearly distinguishable.

2). coherent two-photon illumination
\begin{equation}\label{Image_2}
G_{img2}^{(2)}({\bf{\rho}}_1;{\bf{\rho}}_2)   \propto G_{bg2} + \int  O^*({\bf{\rho}}_{01})   O({\bf{\rho}}_{03}) e^{ik\frac{{\bf{\rho}}^2_{03}-{\bf{\rho}}^2_{01}}{2z_1}}
   \text{somb}^2 \Big(\frac{kD}{2z_1}\big|\pm{\bf{\rho}}_{01} \pm {\bf{\rho}}_{03} - \frac{\Delta{\bf{\rho}}_{21}}{m}\big|\Big)
   d{\bf{\rho}}_{01} d{\bf{\rho}}_{03}  \,.
\end{equation}
where the plus-minus signs mean the total result is the sum of four possible results.
The first term
 $G_{bg2}= \int  O^*({\bf{\rho}}_{01})   O({\bf{\rho}}_{03}) e^{ik\frac{{\bf{\rho}}^2_{03}-{\bf{\rho}}^2_{01}}{2z_1}} \,\,
   \text{somb}^2 \big(\frac{kD|\pm{\bf{\rho}}_{01}\pm{\bf{\rho}}_{03}|}{2z_1}\big)
   d{\bf{\rho}}_{01} d{\bf{\rho}}_{03}$, which is a constant background.
Besides of this constant background, the other term in Eq.~(\ref{Image_2}) shows that interference between pairs of object points $({\bf{\rho}}_{01},{\bf{\rho}}_{03})$ that meet the condition $\pm{\bf{\rho}}_{01} \pm {\bf{\rho}}_{03}-\Delta{\bf{\rho}}_{21}/m=0$ will together contribute to the image spot at position $\Delta{\bf{\rho}}_{21}$ in the image plane. From Eq.~(\ref{Image_2}), we find there are two sets of interference pattern in the image plane caused by two-photon interference:

i). Difference cross-interference pattern when  ${\bf{\rho}}_{01} - {\bf{\rho}}_{03} \pm \Delta{\bf{\rho}}_{21}/m=0$ is satisfied, and the interference term is expressed as
\begin{equation}\label{Diff_img}
\Delta G_{img2}^{(2)}({\bf{\rho}}_1;{\bf{\rho}}_2)   \propto  \int  O^*({\bf{\rho}}_{01})   O({\bf{\rho}}_{03}) e^{ik\frac{{\bf{\rho}}^2_{03}-{\bf{\rho}}^2_{01}}{2z_1}}
   \text{somb}^2 \Big(\frac{kD}{2z_1} \big|({\bf{\rho}}_{03}-{\bf{\rho}}_{01}) \pm \frac{\Delta{\bf{\rho}}_{21}}{m}\big|\Big)
   d{\bf{\rho}}_{01} d{\bf{\rho}}_{03}  \,,
\end{equation}

ii). Sum cross-interference pattern when  ${\bf{\rho}}_{01} + {\bf{\rho}}_{03} \pm \Delta{\bf{\rho}}_{21}/m=0$ is satisfied, and the interference term is expressed as
\begin{equation}\label{Sum_img}
\Delta G_{img2}^{(2)}({\bf{\rho}}_1;{\bf{\rho}}_2)   \propto  \int  O^*({\bf{\rho}}_{01})   O({\bf{\rho}}_{03}) e^{ik\frac{{\bf{\rho}}^2_{03}-{\bf{\rho}}^2_{01}}{2z_1}}
   \text{somb}^2 \Big(\frac{kD}{2z_1} \big|({\bf{\rho}}_{03}+{\bf{\rho}}_{01}) \pm \frac{\Delta{\bf{\rho}}_{21}}{m}\big|\Big)
   d{\bf{\rho}}_{01} d{\bf{\rho}}_{03}  \,,
\end{equation}

\begin{figure}[!htb]
    \centering
    \includegraphics[width=0.8\textwidth]{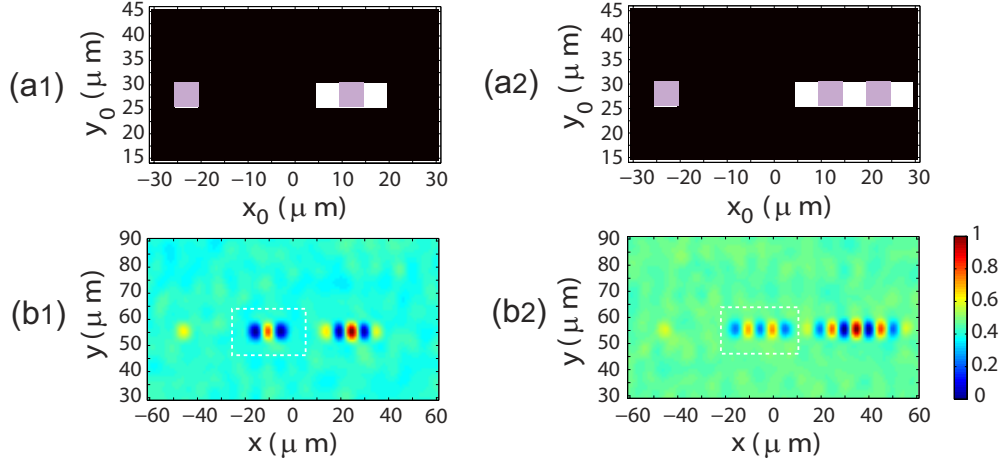}
    \caption{Two-photon imaging of pure-phase objects.
    (a1) and (a2): Pure-phase objects to be imaged. In both cases, the reference point is shown on the left, and a pure-phase object with $\pi$ phase points in a zero phase background is shown on the right. The phase of the reference point is set to be $\pi$ relative to the background phase of the pure-phase object.
    (b1) and (b2): Coherent two-photon imaging corresponding to the pure-phase objects in (a1) and (a2), respectively. With the help of the reference point, a point-to-point correlated image of the pure-phase object is found in the middle marked by a dashed-line box.
    Simulation parameters are the same as those in Fig.~\ref{Fig3_Super_resolution}.
   All plots are normalized so that the minimum and maximum value are 0 and 1, respectively. \label{Fig4_ImagePurePhase}}
\end{figure}

In both sets, despite the light field is totally distorted by the dynamic random medium, the phase information of the object is still preserved in the interference pattern. By using this remarkable feature, we are able to image a pure-phase object by considering the sum cross-interference pattern expressed as Eq.~(\ref{Sum_img}), as we will show in the following.

For the sum cross-interference pattern, there are two kinds of interference spots:
the image spots with point-to-point correlation to the object points when ${\bf{\rho}}_{01}={\bf{\rho}}_{03}$, and a group of additional spots at the middle position of each pair of imaged spots when ${\bf{\rho}}_{01} \neq {\bf{\rho}}_{03}$.
Our method is to introduce a reference point far away from the pure-phase object, and then a point-to-point direct image of the pure-phase object will be constructed through cross interference between the the reference point and the object.
To clearly show this result, we did simulation for a pure-phase object as shown in Fig.~\ref{Fig4_ImagePurePhase} (a1), where the pure-phase object is on the right side while the reference square is put on the left side.
In our simulation, we set the pure-phase object in Fig.~\ref{Fig4_ImagePurePhase} (a1) to be a square of $\pi$ phase shift in its center with respect to the background phase, meanwhile, we set the phase of the reference point to be $\pi$. Figure~\ref{Fig4_ImagePurePhase} (b1) shows the simulation result of sum cross-interference pattern (the difference cross-interference pattern is not shown here, which is spatially distinguishable from the sum cross-interference pattern in the simulation results). In the region between the reference point and the pure-phase object as marked by dashed-line box in Fig.~\ref{Fig4_ImagePurePhase} (b1), we clearly see a direct image of the pure-phase object due to the interference between the reference point and the pure-phase object. Besides, we did simulation with another pure-phase object of two $\pi$ phase squares as shown in Figs.~\ref{Fig4_ImagePurePhase} (a2) and (b2), confirming the ability of our two-photon imaging scheme to image a pure-phase object.

\vspace{12pt}
\noindent \textbf{\large{Discussion}}

In summary, we designed a two-photon imaging scheme, in which a commonly used imaging lens is replaced by a dynamic random medium. Instead of destroying imaging process, the dynamic random medium in our scheme plays a key role for building constructive two-photon interference, thereby, leads to optical two-photon imaging.

With incoherent two-photon illumination, our imaging scheme enables us to obtain a point-to-point direct image of an object, and the imaging resolution surpasses the well known Rayleigh resolution bound by a factor of 2.
Note that this resolution enhancement was achieved in the two-photon diffraction or interferometer schemes perviously~\cite{fonseca1999measurement,boto2000quantum,edamatsu2002measurement,d2001two,scarcelli2004two,wang2004subwavelength,xiong2005experimental,hong2013subwavelength,hong2015super}, but not in a point-to-point direct imaging scheme.
Recent advances show that, by introducing two-photon correlation in lens-assisted imaging schemes, the resolution is enhanced by a factor of $\sqrt{2}$~\cite{dertinger2009fast,guerrieri2010sub,oh2013sub,schwartz2012improved,schwartz2013superresolution,monticone2014beating,xu2015experimental}.
Our two-photon imaging scheme breaks the Rayleigh resolution bound by a factor of 2, reaching the fundamental Heisenberg limit based on interference of multi-photon de Broglie waves~\cite{jacobson1995photonic,giovannetti2009sub}.

Besides, with coherent two-photon illumination, the phase information of the object is preserved in two-photon interference patterns of the totally distorted light propagating through a dynamic random medium. By introducing a reference point, we show that direct image of a pure-phase object can be obtained in our two-photon imaging scheme.

Our two-photon imaging scheme assisted by a dynamic random medium is practically achievable by using current technology. Since our two-photon imaging scheme does not rely on finely designed lens for imaging, it may have application in building high-resolution microscope with new features such as long working distance. These results show the potential benefits of two-photon or multi-photon imaging in complementary to current widely used single-photon imaging system.

\vspace{36pt}
\noindent \textbf{\large{Methods}}

We numerically calculated the two-photon amplitude in the image plane by using the Fresnel diffraction formula, and then obtained the two-photon correlation function. To get an enough ensemble average with respect to the aperture function ${\cal{R}}({\bf{\rho}}_r,t)$ of the dynamic random medium, we repeated the calculation of the two-photon correlation function 10 000 times each with a different random phasor ${\cal{R}}_i({\bf{\rho}}_r) \, (i=1,2,\cdots, 10\, 000)$. Consequently, an ensemble-averaged two-photon correlation function is obtained as the final result.

\vspace{12pt}


\vspace{24pt}
\noindent \textbf{\large{Acknowledgement}}

The author thanks Yu Wang for helpful discussions. This work is supported by the National Natural Science Foundation of China (Grant No. 11604150).

\end{document}